\documentclass{WileyMSP-template}

\begin{document}

\pagestyle{fancy}
\rhead{\includegraphics[width=2.5cm]{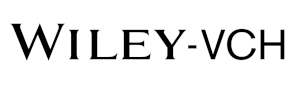}}

\title{Modification of the magnetic and electronic properties of the \newline graphene-Ni(111) interface via halogens intercalation}

\maketitle

\author{Yong Zhou},
\author{Roman Ovcharenko},
\author{Beate Paulus},
\author{Yuriy Dedkov}, 
\author{Elena Voloshina}

\begin{affiliations}
Y. Zhou, Prof. Y. Dedkov, Prof. E. Voloshina\\
Department of Physics, Shanghai University, 200444 Shanghai, China\\

\medskip

Dr. R. Ovcharenko\\
Max-Born-Institut f\"ur Nichtlineare Optik und Kurzzeitspektroskopie, 12489 Berlin, Germany

\medskip

Prof. B. Paulus, Prof. E. Voloshina\\
Institut f\"ur Chemie und Biochemie, Freie Universit\"at Berlin, 14195 Berlin, Germany

\medskip

Prof. Y. Dedkov, Prof. E. Voloshina\\
Centre of Excellence ENSEMBLE3 Sp. z o.\,o., ul. Wolczynska 133, 01-919 Warsaw, Poland\\
Email Address: yuriy.dedkov@icloud.com; elena.voloshina@icloud.com

\end{affiliations}


\medskip

\begin{abstract}

Electronic decoupling of graphene from metallic and semiconducting substrates via intercalation of different species is one of the widely used approaches in studies of graphene. In the present work the modification of the electronic and magnetic properties of graphene on ferromagnetic Ni(111) layer via intercalation of halogen atoms (X\,=\,F, Cl, Br) is studied using the state-of-the-art density-functional theory approach. It is found that in all gr/X/Ni(111) intercalation systems a graphene layer is fully electronically decoupled from the ferromagnetic substrate; however, different kind (electron or hole) and level of doping can be achieved. Despite the extremely small magnetic moment of C-atoms in graphene observed after halogens intercalation, the sizeable spin-splitting up to $35$\,meV for the linearly dispersing graphene $\pi$ bands is found. The obtained theoretical data bring new ideas on the formation of the graphene-ferromagnet interfaces where spin polarized free-standing graphene layer can be formed with the possible application of these systems in electronics and spintronics. 

\end{abstract}

\section{Introduction}

Graphene (gr), a single 2D layer formed by carbon atoms, still attracts a lot of attention of many scientists and engineers~\cite{Geim.2009,Neto.2009,Sarma.2011,Kumar.2021}. Despite the absence of the big breakthroughs in the production of the graphene-based serial products as well as the scepticism on the future commercialisation of the graphene large-scale production~\cite{Peplow.2015,Park.2016o1}, the unique electronic properties of graphene stimulate further studies allowing to understand many fundamental physical phenomena. Among recent discoveries are the formation of the graphene-hBN superlattices with ``Hofstadter butterfly'' spectrum for massive fermions~\cite{Yankowitz.2012,Hunt.2013,Ponomarenko.2013,Yankowitz.2019} and observation of the strongly correlated state and superconductivity in small-angle twisted bilayer graphene~\cite{Cao.2018,Cao.2018mjda,Andrei.2020}.

In many works graphene is considered and studied as not free-standing, but supported by different substrates. Here, the most interesting objects are graphene layers epitaxially grown on different supports, semiconducting or metallic~\cite{Wang.2013e6c,Lee.2014,Dedkov.2020,Batzill.2012,Dedkov.2015}. In the later case many exciting phenomena were found, like, for example, protective properties of a single graphene layer on metals~\cite{Dedkov.2008xen,Dedkov.2008kg,Sutter.20109b,Borca.2009} and formation of the graphene-metal moir\'e structures of different periodicities~\cite{Coraux.2008,Pletikosic.2009,Voloshina.2012py,Stradi.2013,Voloshina.2016iy8}. One of such property for graphene-metal interfaces is the possibility to realize the perfect spin-filtering effect in the FM/$n$-ML-gr/FM sandwiches, where FM is the close-packed surface of ferromagnetic material, Ni(111) or Co(0001)~\cite{Karpan.2007,Karpan.2008}. This effect is explained by the effective overlapping of the valence band states of graphene and FM material around the $K$ point of hexagonal Brillouin zone only for one spin channel, with the significant decrease of the conductance for the other spin channel. However, such overlap of valence states leads to the appearance of the non-zero spin polarization of the graphene-derived $\pi$ states, that was confirmed experimentally~\cite{Weser.2010,Weser.2011,Marchenko.2015,Usachov.2015js7}. The spin-filtering effect has a significant value for $n>2$, and the hybridization of the electronic states of graphene and FM at the interface leads to the very small value of magnetoresistance ratio for the case $n=1$~\cite{Karpan.2007}. Many ways were proposed in order to decouple the electronic states of graphene and FM, and to increase the spin-filtering effect for the case of a sandwiched single graphene layer, like the insertion of an hBN monolayer or layers of $s,p$-materials~\cite{Karpan.2007,Karpan.2008,Voloshina.2011}. However, in such systems the intercalation of hBN or noble metals (NMs) between graphene and FM material can lead to the collapse of magnetic moment of interface atoms and finally to the absence of the spin-filtering effect. Moreover, the formation of sharp interfaces in the case of noble metal intercalation is a challenging task due to the fast and uncontrollable alloy formation at the NM-FM interface.

Here, we present a new way to decouple graphene from Ni(111) ferromagnetic surface using the halogens atoms intercalation (X: F, Cl, Br). The possible ways of intercalation of these atoms in the gr/FM interfaces are discussed. It is found that, contrary to the previously discussed case of gr/Co interface, where spin-polarized electrons of mainly $d$-character are found at the $K$ point with significantly reduced Fermi velocity ($v_F$), in the present gr/X/FM systems the spin-polarized electrons have pure graphene-$\pi$ character with $v_F\approx1\times10^6\,\mathrm{m/s}$. Graphene in the gr/X/FM systems have different level of the electron- and hole-doping depending on the type and concentration of the intercalated halogen atoms. However, in all cases graphene-derived $\pi$ states around the $K$ point and around the Fermi level ($E_F$) are spin-polarized with a spin splitting reaching $35$\,meV. Additionally to the analysis of the band structure effects for the studied systems, we performed simulations of the C\,$K$ near-edge x-ray absorption fine structure (NEXAFS) spectra, which can be used as a reference for the comparison with experimental ones in future works. Our theoretical findings can be of interest for the future spectroscopic experiments of the discussed systems and also in the studies of the spin-transport phenomena in the gr/FM systems and for the discussion of the possible realization of FM/gr/FM spin-filter.

\section{Results and discussion}

In the present study halogen atoms (F, Cl, Br) of different concentrations were placed at the gr/Ni(111) interface with the goal to study the decoupling of the electronic states of a graphene layer from those of ferromagnetic substrate and to preserve the spin splitting of graphene-derived $\pi$ states. The possibility of intercalation of halogen atoms in the gr/SiC and gr/Ir interfaces was discussed and realized experimentally in Refs.~\cite{Walter.2011} and \cite{Vinogradov.2012c1}. In the first case, F-atoms were produced via reaction of XeF$_2$ and Mo and then intercalated in the gr/SiC interface. The resulting gr/F/SiC(0001) system is characterized by the presence of decoupled free-standing $p$-doped graphene with the position of the Dirac point ($E_D$) at $+0.79$\,eV above $E_F$~\cite{Walter.2011}. In case of the Cl-atoms intercalation in gr/Ir(111), it is achieved via thermal decomposition of the pre-deposited FeCl$_3$ thin layer. Such treatment leads to the formation of the gr/Cl/Ir(111) system where graphene is also $p$-doped and $E_D-E_F=+0.6$\,eV~\cite{Vinogradov.2012c1}. These works suggest the feasibility of the gr/X/Ni(111) systems formation, which electronic and magnetic properties are investigated in the present study. 

\begin{figure}
\center
\includegraphics[width=1.0\textwidth]{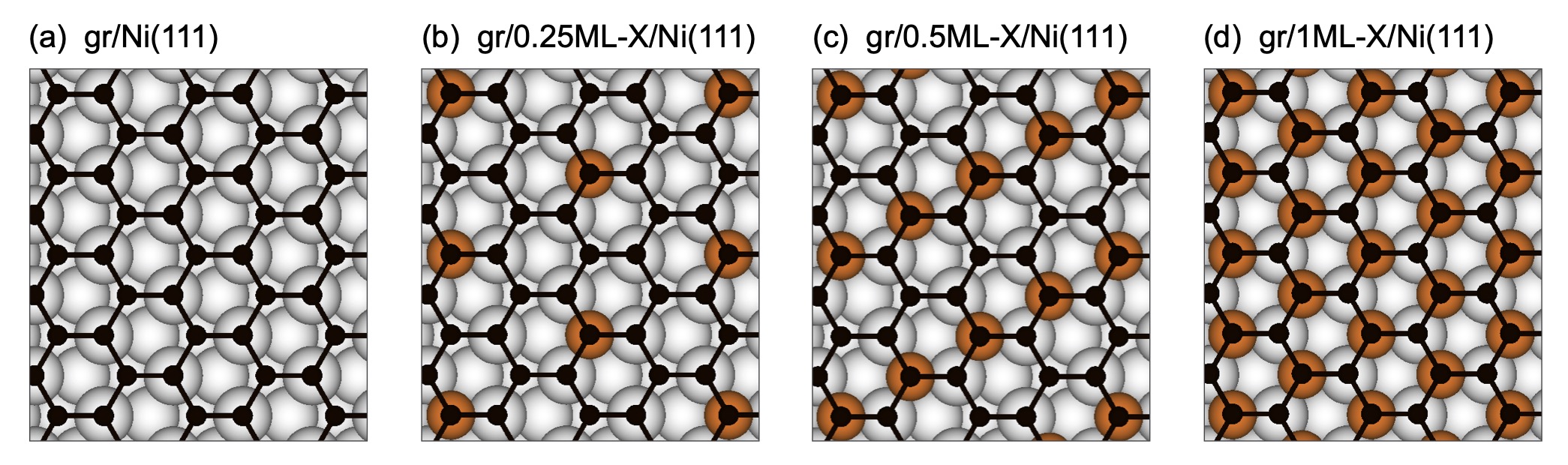}\\
\caption{Top views of different X-based intercalation structures: (a) parent graphene/Ni(111),
(b) graphene/0.25ML-X/Ni(111), (c) graphene/0.5ML-X/Ni(111), and (d)  graphene/1ML-X/Ni(111). Spheres of different size and colour represent atoms of different types:  Grey, brown, and black spheres represent Ni, X, and C atoms, respectively. }
\label{fig:Fig_structures}
\end{figure}

The crystallographic models for the gr/Ni(111) and gr/X/Ni(111) interfaces with concentrations of halogen atoms $0.25$\,ML, $0.5$\,ML and $1$\,ML are shown in Figure~\ref{fig:Fig_structures}. Here, the concentration of $1$\,ML corresponds to one halogen atom per one unit cell (or one carbon ring) of graphene. As was shown in a series of previous works and also confirmed by the present results, a single graphene layer is adsorbed on Ni(111) in the so-called \textit{top-fcc} configuration, where one carbon atom of the graphene unit cell is adsorbed above Ni interface atom and second carbon atom is located at the \textit{fcc} hollow site of the Ni(111) surface. Such geometry leads to the small gr-Ni(S) [Ni(S) denotes the topmost Ni-layer] distance of $2.086$\,\AA\ with the adsorption energy for graphene on Ni(111) equal to $-159$\,meV/C-atom (see Table~\ref{tab:results}), giving excellent agreement with previous theoretical results as well as with available experimental data~\cite{Dedkov.2010,Weser.2011,Voloshina.2011,Dahal.2014,Gamo.1997,Parreiras.2014}. As was discussed earlier, the adsorption of graphene on Ni(111) leads to the $n$-doping of graphene with downward shift of the graphene-derived $\pi$ states, followed by the strong hybridization of the graphene-$\pi$ and Ni-$3d$ valence band states, followed by the formation of hybrid states in the vicinity of $E_F$ with the predominant Ni-$3d$ character. As a result the magnetic moment on carbon atoms appears, that was confirmed by present and previous theoretical calculations as well as in spectroscopic measurements~\cite{Bertoni.2005,Weser.2010,Matsumoto.2013}. At the same time the magnetic moment of the Ni(S) atoms is strongly reduced to $0.505\,\mu_B$ (Table~\ref{tab:results}) compared to the values of $0.71\,\mu_B$ and $0.636\,\mu_B$ for the clean Ni(111) surface and Ni-bulk, respectively. 

\begin{figure}
\center
\includegraphics[width=\textwidth]{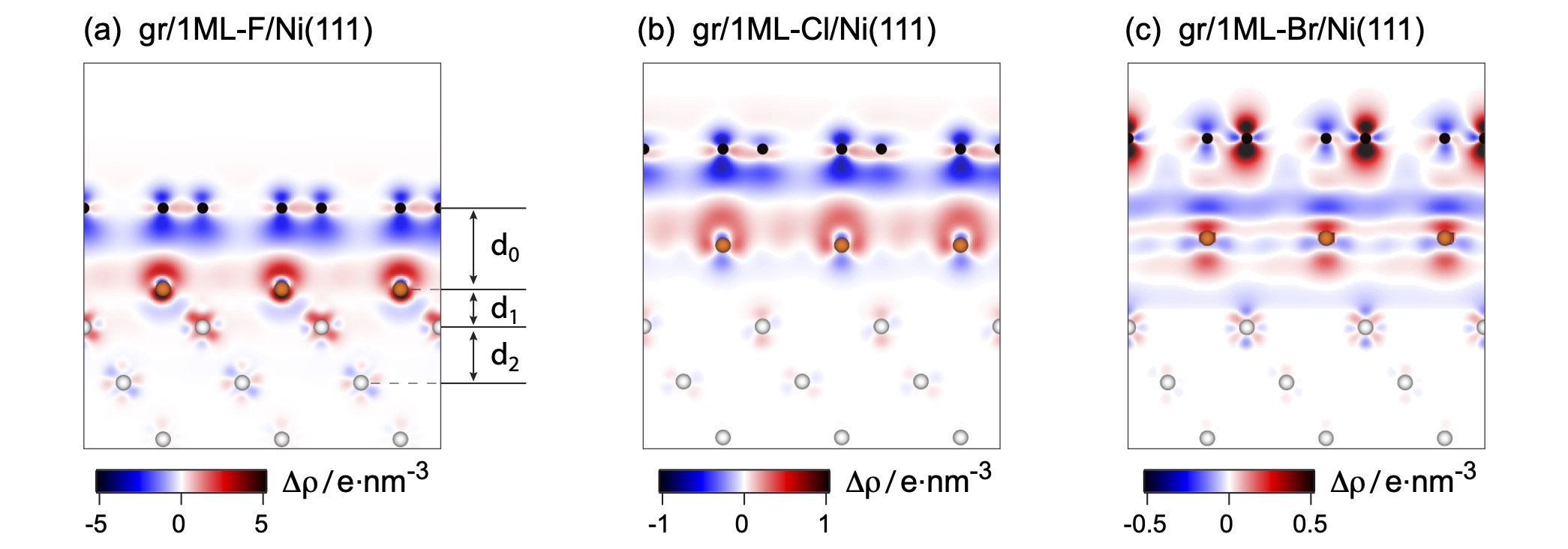}\\
\vspace{1cm}
\caption{Side views of gr/1ML-X/Ni(111) intercalation structures: (a) X = F, (b) X = Cl, (c) X = Br, taken along the graphene arm-chair edge and overlaid with electron charge difference maps $\Delta\rho(r) = \rho_\mathrm{gr/sub}(r) - [\rho_\mathrm{gr}(r) + \rho_\mathrm{sub}(r)]$ (gr: graphene; sub: substrate). The distances between the graphene overlayer and the substrate layer, between the interface substrate layer and the second substrate layer, and between the second and third substrate layers are marked with $d_0$, $d_1$ and $d_2$, respectively. }
\label{fig:Fig_chg}
\end{figure}

The intercalation of halogen atoms (F, Cl and Br) in gr/Ni(111) leads to the dramatic modifications of the electronic and magnetic properties of graphene and interface. According to our density functional theory (DFT) calculations, the most energetically favourable configuration for the gr/X/Ni(111) system of different halogen concentrations is when halogen atom is placed in the \textit{fcc} hollow site of Ni(111) under C$^{fcc}$ atom of a graphene unit cell. The distances between graphene and intercalated halogen ($d_0$ in Tab.~\ref{tab:results}, see also Fig.~\ref{fig:Fig_chg}) and those between intercalated halogen and underlying Ni ($d_1$ in Tab.~\ref{tab:results}, see also Fig.~\ref{fig:Fig_chg}) increase with growing radius in the row F -- Cl -- Br. Obviously, the increase of $d_0$ indicates the decoupling of 2D layer from the substrate. Furthermore, the correlation of the growing $d_0$ and $d_1$ with modification of the doping level of a graphene layer (see the discussion below), emphasises the importance of the electrostatic interaction in the considered systems. The latter is quite strong in the case of the gr--F--Ni and gr--Cl--Ni systems, where significant charge transfers between involved layers are observed. The calculated adsorption energy in these cases are $-98$ \,meV/C-atom, $-41$\,meV/C-atom, and $-55$\,meV/C-atom, for X = F, Cl, Br respectively, which support the previous statements and confirm the very weak interaction for the gr/X/Ni(111) systems.

\begin{figure}
\center
\includegraphics[width=\textwidth]{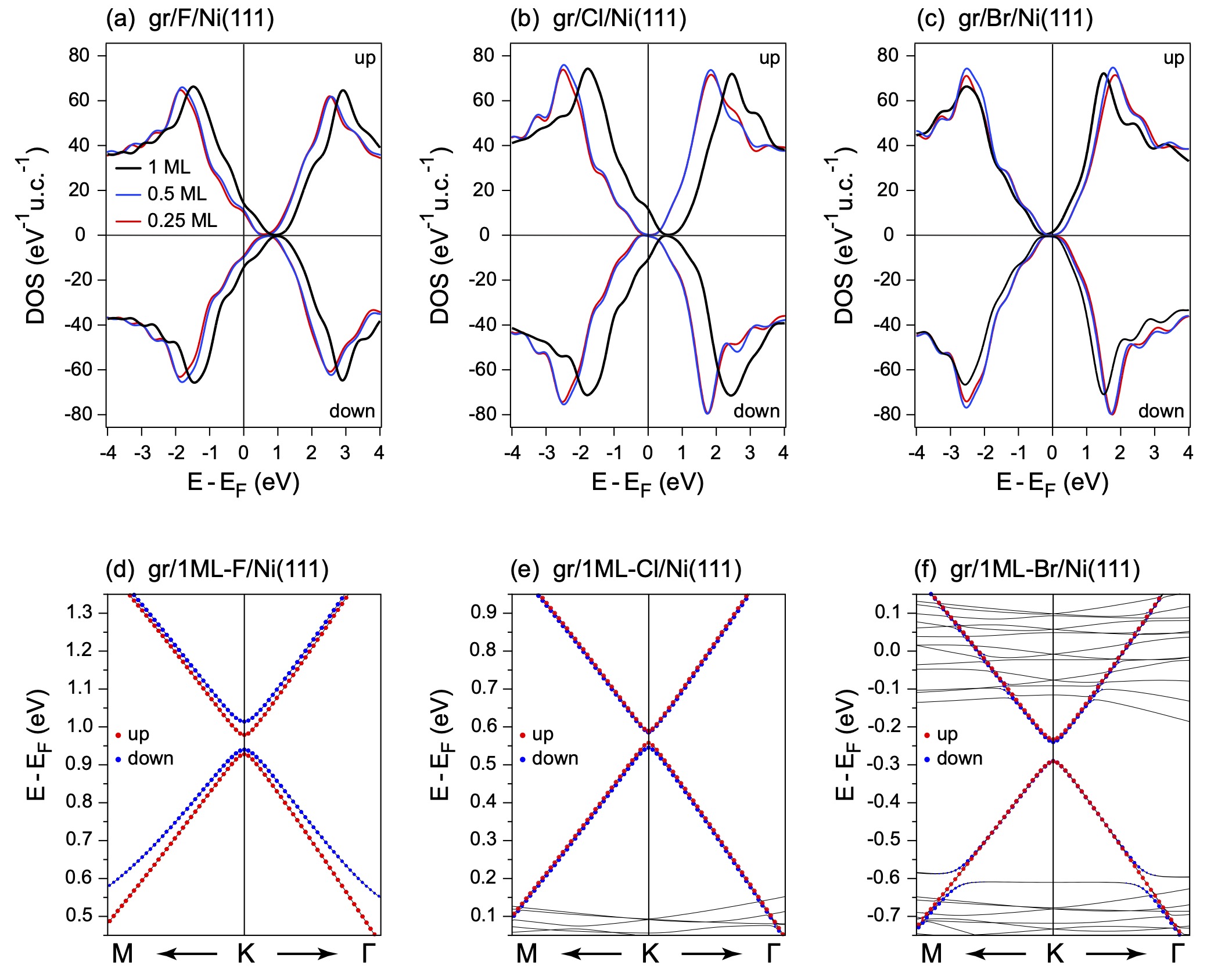}\\
\caption{(upper panel) The C-atom projected density of states for (a) the graphene/F/Ni(111),  (b) graphene/Cl/Ni(111), (c) graphene/Br/Ni(111) systems. (lower panel) Calculated electron energy dispersion of the graphene-derived $\pi$ valence band states along the main directions of the hexagonal Brillouin zone for the graphene/1ML-X/Ni(111) system with (d) X = F, (e) X= Cl, (f) X = Br. } 
\label{fig:Fig_dos+bands}
\end{figure}

Figure~\ref{fig:Fig_dos+bands} (upper panel) shows the calculated C-atom-projected density of states (DOS) in the vicinity of $E_F$ for gr/X/Ni(111) with different X-atom concentration. In case of the gr/F/Ni(111) and gr/Cl/Ni(111) interfaces, graphene is $p$-doped and doping level is increased with increasing the concentration of the X-atoms, giving positions of the Dirac point at $E_D-E_F=+0.965$\,eV and $E_D-E_F=+0.569$\,eV for $1$\,ML of intercalated F and Cl, respectively (Fig.~\ref{fig:Fig_dos+bands}(a,b)). In case of intercalated Br, graphene is slightly $n$-doped, with a similar trend of increasing the doping level with growing the concentration of Br-atoms and position of the Dirac point reaches $E_D-E_F=-0.254$\,eV for gr/$1$-ML\,Br/Ni(111) (Fig.~\ref{fig:Fig_dos+bands}(c)). 

The doping level of graphene and charge distribution at the systems' interfaces are nicely visualised in electron density differences maps. Due to the high electronegativity of F-atoms, they strongly attract electron density from C- and Ni-atoms (Fig.~\ref{fig:Fig_chg}(a)). Also, the relatively strong chemical bonds are formed between F and Ni. Such charge redistribution dramatically reduces the \textit{screening} of magnetic moments of Ni(S) atoms (Ni\,$4s,p$ electrons' density is spatially shifted towards F-atoms) that leads to the significant increase of magnetic moments of these atoms to $\approx1.05\,\mu_B$ compared to $0.636\,\mu_B$ for Ni-bulk (see Table~\ref{tab:results}). The similar effect, although quantitatively reduced, is also observed for the gr/Cl/Ni(111) and gr/Br/Ni(111) systems; here, the magnetic moment of Ni(S) atoms reaches only $0.688\,\mu_B$ and $0.646\,\mu_B$, respectively, compared to $0.505\,\mu_B$ calculated for Ni(S) in gr/Ni(111). In all considered systems the total induced magnetic moment of C-atoms is significantly reduced to the value of $0.002\,\mu_B$ at maximum pointing out that halogen atoms effectively screen the magnetic interaction at the interface.

Although the previously described behaviour of a graphene layer in the gr/X/Ni(111) systems indicates its free-standing character, the graphene electronic structure around the $K$ point is significantly modified. Firstly, a sizeable band gap is opened directly at the Dirac point with a value of $63$\,meV, $38$\,meV, and $57$\,meV for X\,=\,F, Cl, and Br, respectively, for a concentration of halogen atoms corresponding to $1$\,ML (Fig.~\ref{fig:Fig_dos+bands} (lower panel); see also Fig.\,S1 of Supporting Information for band structures presented in a wider energy range). This effect can obviously be assigned to the broken sub-lattice symmetry in a graphene layer due to the intercalation of halogen atoms and formation of different electrostatic potentials for two carbon atoms (Fig.~\ref{fig:Fig_chg}(a-c)). The most interesting and intriguing fact obtained from the analysis of the electronic structure in the vicinity of the K point for the gr/X/Ni(111) systems is the observation of the spin-splitting for the graphene-derived $\pi$ bands. Although this effect is quite small in case of X\,=\,Cl and Br, for the gr/F/Ni(111) system this splitting reaches $12$\,meV and $35$\,meV at the K point for the $\pi$-band branches below and above the Dirac point, respectively. A possible reason for this splitting could be the interplay of different factors observed for the gr/F/Ni(111) interface, like the increased magnetic moment of Ni(S) atoms, existence of the large gradient of the electrostatic potential at the interface, large charge transfers, etc.

The modification of the electronic structure of graphene on metallic substrates upon intercalation of different species can be effectively monitored using different electron spectroscopy methods, like x-ray photoelectron spectroscopy (XPS) or NEXAFS spectroscopy. In the first case the chemical state of carbon atoms is monitored allowing to get information about graphene's doping level. In the case of NEXAFS spectroscopy at the C\,$K$ absorption edge of graphene the $1s\rightarrow\pi^*,\sigma^*$ transitions are used to probe the unoccupied valence band states above the Fermi level and to get information about orbital character and spatial localization of these states (the so-called \textit{search-light-like} effect), doping level of graphene, strength of the graphene-substrate interaction, etc~\cite{Wessely.2005,Wessely.2006,Ovcharenko.2013,Voloshina.20136kp}.

\begin{figure}
\center
\includegraphics[width=0.5\textwidth]{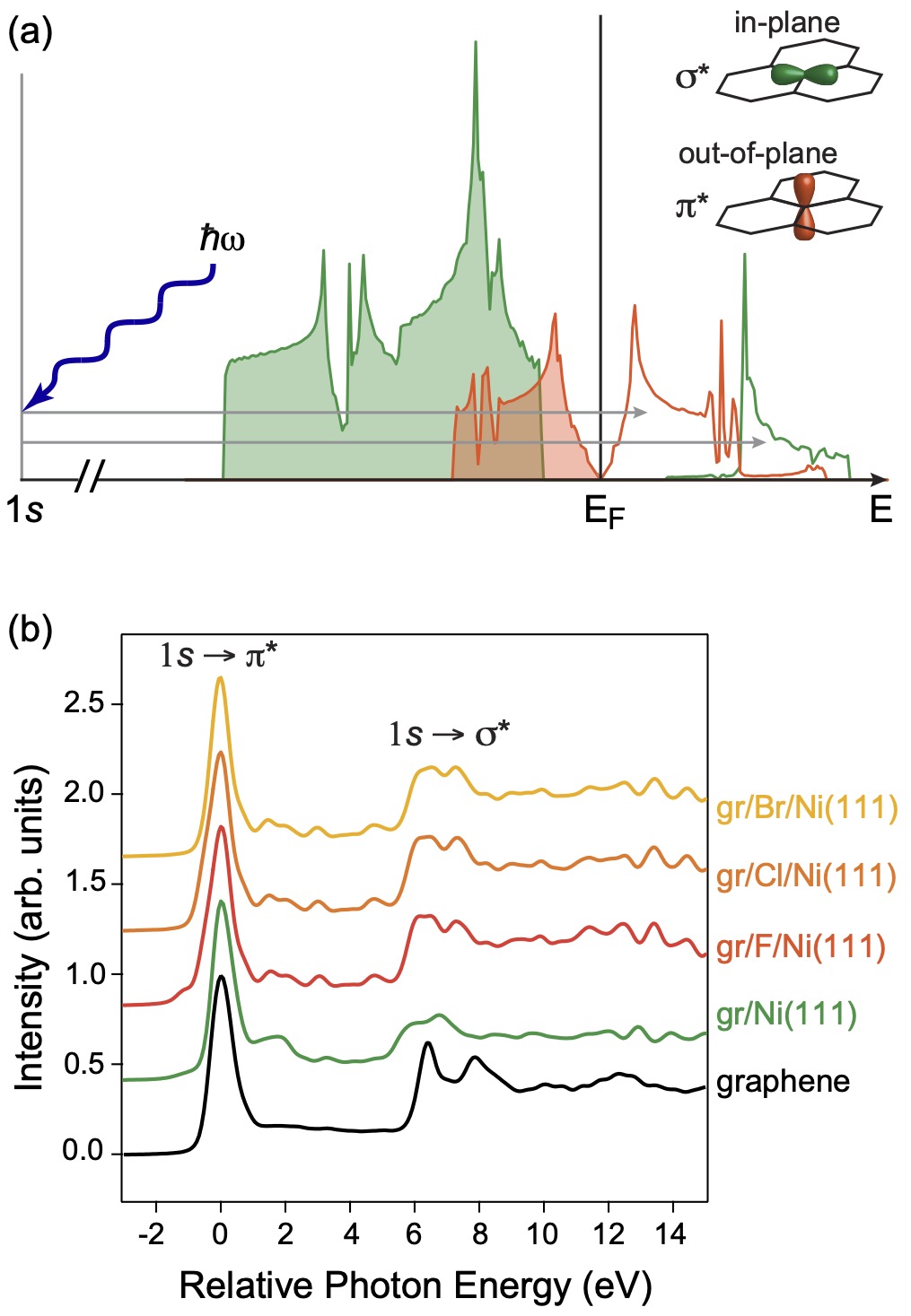}\\
\caption{(a) The scheme of the X-ray light absorption process. The light from the synchrotron light source is scanned around the energy corresponding to the binding energy of the particular core level and in this case the absorption coefficient is proportional to the density of the valence band states above $E_F$. The orbital selectivity is reached via relative orientation of the linearly polarised light and the sample. (b) Calculated C $K$ NEXAFS spectra for $\alpha = 40^\circ$ for graphene, graphene/Ni(111), graphene/1ML-F/Ni(111), graphene/1ML-Cl/Ni(111), and graphene/1ML-Br/Ni(111). $\alpha$ is an angle between electric ﬁeld vector of x-ray and the normal to the surface.}
\label{fig:Fig_nexafs}
\end{figure}

Using theoretical approach discussed in Refs.~\cite{Ovcharenko.2013} and \cite{Voloshina.20136kp} the C\,$K$ NEXAFS spectra of the gr/X/Ni(111) systems were calculated (Fig.~\ref{fig:Fig_nexafs}), where all NEXAFS spectra are presented in the relative photon energy scale with the position of the main peak for the C $1s\rightarrow\pi^*$ transition at $0$\,eV. The reference spectra for the free-standing graphene and gr/Ni(111) are in very good agreement with previously calculated NEXAFS spectra for these objects as well as with the respective experimental data~\cite{Weser.2010,Rusz.2010,Ovcharenko.2013,Voloshina.20136kp}. The NEXAFS intensity in the approximate ranges $-2...+3$\,eV and $+5...+11$\,eV corresponds to the C $1s\rightarrow\pi^*$ and C $1s\rightarrow\sigma^*$ transitions, respectively, of the core electron from $1s$ level on the corresponding unoccupied valence band states above $E_F$. After adsorption of graphene on Ni(111), the relatively \textit{strong} interaction at the interface and the respective redistribution of the valence band states leads to the strong modification of the NEXAFS spectra: (i) additional peak appears at $\approx2$\,eV, which can be assigned to the respective predominant contributions of two different carbon atoms in gr/Ni(111) to the corresponding unoccupied valence states above $E_F$ and (ii) the energy difference between the C $1s\rightarrow\pi^*$ and C $1s\rightarrow\sigma^*$ transitions, with the significant shape modifications of both peaks, is decreased by $\approx0.5$\,eV, that can be assigned to the partial $sp^2$-to-$sp^3$ rehybridization due to adsorption of graphene on Ni(111)~\cite{Weser.2010,Rusz.2010}.

The intercalation of halogen atoms in gr/Ni(111) and formation of the respective interfaces leads to the restoring of the C\,$K$ NEXAFS spectra (Fig.~\ref{fig:Fig_nexafs}). Firstly, the respective splitting of the peak corresponding to the C $1s\rightarrow\pi^*$ is much weaker now (the observed oscillations of the intensity can be assigned to the computational artefacts due to the size of the unit cell used in the calculations). Also, the energy splitting between the C $1s\rightarrow\pi^*$ and C $1s\rightarrow\sigma^*$ transitions is restored and its value of $\approx6.5$\,eV is the same as for a free-standing graphene layer. These facts reflect the free-standing character of the graphene valence band states for the gr/X/Ni(111) systems, where interaction with substrate is significantly reduced compared to gr/Ni(111). Additionally to the above described changes, a pronounced step-like pre-edge feature (in the relative photon energy range of $-1.5\dots-0.9$\,eV) is observed in the C\,$K$ NEXAFS spectrum of gr/F/Ni(111). This immediately indicates the lowering of the Fermi level upon intercalation of F-atoms in gr/Ni(111) and the position of this peak ($-1.08$\,eV) correlates with the respective doping level of graphene for this system. The similar effect of the lowering of the Fermi level after the Cl-atoms intercalation, but not so pronounced, is also found for the gr/Cl/Ni(111) system - here low-energy shoulder is clearly observed for the C $1s\rightarrow\pi^*$ peak. In the case of the gr/Br/Ni(111) system, the small low-energy off-set (of approximately $75$\,meV) of the NEXAFS intensity indicates the free-standing weakly $n$-doped graphene. The presented theoretical results for the simulation of the C\,$K$ NEXAFS spectra of the graphene-metal interfaces upon halogen-atom intercalation are in very good agreement for the available experimental data for the gr/Cl/Ir(111) system (shape of spectra, value of the energy splitting between the C $1s\rightarrow\pi^*$ and C $1s\rightarrow\sigma^*$ transitions, appearance of a pronounced step-like pre-edge feature)~\cite{Vinogradov.2012c1}, confirming correctness of the presented approach for the description of the gr/X/Ni(111) systems. 

\section{Conclusions}

In conclusion, we theoretically investigate the modification of the electronic and magnetic properties of the gr/Ni(111) interface upon intercalation of halogen atoms: F, Cl, and Br. It is found that in case of $1$\,ML of intercalated F and Cl, the strong $p$-doping of graphene is developed with the increase of the halogen-atoms concentration ($E_D-E_F=+0.965$\,eV and $E_D-E_F=+0.569$\,eV, respectively), whereas $1$\,ML of intercalated Br-atoms keep a graphene layer $n$-doped ($E_D-E_F=-0.254$\,eV). In all intercalation systems, graphene behaves as a free-standing layer with its electronic states effectively decoupled from the valence band states of the substrate. Despite this fact, the distance between graphene and halogen-atoms layers is strongly varied from $2.916$\,\AA\ for gr/F/Ni(111) to $3.605$\,\AA\ for gr/Br/Ni(111) indicating the importance of the electrostatic forces in the interaction of the doped graphene and underlying X/Ni(111). Although, the magnetic moment of C-atoms of graphene is almost fully quenched by the intercalation of halogens, the sizeable spin-splitting is found for the graphene-derived $\pi$ bands around the K point, which can be assigned to the strong increase of the magnetic moment of Ni(S) atoms and strong redistribution of the charge density in the gr/F/Ni(111) interface. For all intercalation gr/X/Ni(111) systems the C\,$K$ NEXAFS spectra were simulated giving very good agreement with the available spectroscopic data for the graphene-metal interfaces. Our theoretical results demonstrate new approach for the decoupling of the electronic states of graphene from the strongly-interacting ferromagnetic substrates allowing to keep the linear dispersion around the K point of the graphene's electronic states together with sizeable spin splitting of the graphene $\pi$ band. Such approach can be useful for the development of the future ideas on the application of graphene-ferromagnet systems in electronics and spintronics. 

\section{Computational details}\label{CompDetails}

The DFT calculations based on plane-wave basis sets of $500$\,eV cutoff energy were performed with the Vienna \textit{ab initio} simulation package~\cite{Kresse.1996,Kresse.1994}. The Perdew-Burke-Ernzerhof exchange-correlation functional~\cite{Perdew.1996} was employed. The electron-ion interaction was described within the projector-augmented wave method~\cite{Blochl.1994gal} with C ($2s$, $2p$), Ni ($3d$, $4s$), F ($2s$, $2p$), Cl ($3s$, $3p$), and Br ($4s$, $4p$) states treated as valence states. The Brillouin zone integration was performed on the $\Gamma$-centred symmetry-reduced Monkhorst-Pack meshes using a Methfessel-Paxton smearing method of first order with $\sigma = 0.15$\,eV, except for the calculation of total energies. For these calculations, the tetrahedron method with Bl\"ochl corrections~\cite{Blochl.1994} was employed. The $k$-mesh for sampling of the supercell Brillouin zone was chosen to be as dense as $24\times24$ when folded up to the simple graphene unit cell. Dispersion interactions were considered adding a $1/r^6$ atom-atom term as parameterised by Grimme (``D2'' parameterisation).~\cite{Grimme.2006}

The studied systems were modelled by symmetric slabs. The used supercells have a ($2\times2$) lateral periodicity with respect to graphene and contain $13$ layers of metal atoms ($4$ atoms per layer) and a graphene sheet ($8$ atoms per layer) adsorbed on both sides of the slab. In the case of the gr/X/Ni(111) system, X-atoms ($1$, $2$ and $4$ atoms in the case of $0.25$\,ML, $0.5$\,ML and $1$\,ML concentrations, respectively) are placed from both sides of the slab between graphene and Ni(111). The lattice constant in the lateral plane was set according to the optimised lattice constant of Ni(111), $a = 2.488$\,\AA. A vacuum gap in all cases was larger than $18$\,\AA. For all these structures the ions of the $9$ middle inner layers were fixed at their bulk positions during the structural optimisation procedure, whereas the positions of all other ions were fully relaxed until forces became smaller than $0.01$\,eV\,\AA$^{-1}$. The convergence criteria for energy was set equal to $10^{-5}$\,eV.  

The NEXAFS spectra simulations were obtained with ELSA-software~\cite{ELSA} according to the procedure described in Refs.~\cite{Ovcharenko.2013} and \cite{Voloshina.20136kp}. To obtain an input for the NEXAFS spectra simulations, we employed large supercells of ($6\times6$) and ($4\times4$) periodicity when studying free-standing graphene and gr/Ni(111) or gr/X/Ni(111) systems, respectively. In the latter case the number of nickel layers was reduced to $5$. 


\bigskip
\textbf{Supporting Information} 
Supporting Information is available from the Wiley Online Library or from the author.

\bigskip

\medskip
\textbf{Acknowledgements} \par 

Y.D. and E.V.  thank the ``ENSEMBLE3 - Centre of Excellence for nanophotonics, advanced materials and novel crystal growth-based technologies'' project (GA No. MAB/2020/14) carried out within the International Research Agendas programme of the Foundation for Polish Science co-financed by the European Union under the European Regional Development Fund and the European Union’s Horizon 2020 research and innovation programme Teaming for Excellence (GA. No. 857543) for support of this work. B.P. acknowledges  funding  from the German Research Foundation (DFG) through the collaborative research center SFB 1349 ``Fluorine-Specific Interactions'', project ID 387284271. The North-German Supercomputing Alliance (HLRN) and the computing facilities (ZEDAT) of the Freie Universit\"at Berlin are acknowledged for providing computer time.

\medskip
\bibliographystyle{MSP}

\clearpage
\begin{table}
\caption{Results obtained for the graphene/Ni(111) interface and for the graphene/1ML-X/Ni(111) (X: F, Cl, Br):  
$d_0$ (in \AA) is the distance between the graphene overlayer and the substrate layer (two values for the two nonequivalent carbon atoms are indicated) (see Fig.~\ref{fig:Fig_chg}); 
$d_1$ (in \AA) is the distance between the interface substrate layer and the second substrate layer (see Fig.~\ref{fig:Fig_chg}); 
$d_2$ (in \AA) is the distance between the second and third substrate layers; $m_\mathrm{C}$ (in $\mu_\mathrm{B}$) is the interface carbon spin magnetic moment (two values for the nonequivalent carbon atoms are indicated) (see Fig.~\ref{fig:Fig_chg}); 
$m_\mathrm{Ni}$ (in $\mu_\mathrm{B}$) is the interface/bulk Ni spin magnetic moment; 
$E_\mathrm{ads}$ (in meV/C-atom) is the adsorption energy, defined as $E_\mathrm{ads} = E_\mathrm{gr/sub} - (E_\mathrm{gr} + E_\mathrm{sub})$, where $E_\mathrm{gr/sub}$ is the total energy of the graphene/substrate system, and $E_\mathrm{gr}$ and $E_\mathrm{sub}$ are the energies of the relaxed fragments; $E_D - E_F$ (in eV) is the position of the Dirac point with respect to the Fermi energy (two values are given for the spin-up and spin-down channels, respectively); $E_\mathrm{g}$ (in meV) is the band gap width (two values are given for the spin-up and spin-down channels, respectively).}
\label{tab:results}
  \begin{tabular}[htbp]{@{}lcccc@{}}
\hline
Property 			& gr/Ni(111) 		& gr/F/Ni(111) 		& gr/Cl/Ni(111)		& gr/Br/Ni(111) \\
\hline
$d_0$ 			&$2.088$/$2.084$	&$2.916$/$2.916$	&$3.482$/$3.482$	&$3.605$/$3.605$\\
$d_1$ 			&$2.023$			&$1.389$			&$2.939$			&$3.234$ \\
$d_2$ 			&$2.019$			&$2.008$			&$2.000$			&$1.998$ \\
$m_\mathrm{C}$ 	&$-0.019$/$0.033$	&$0.002$/$0.000$	&$-0.002$/$0.000$	&$0.000$/$-0.002$\\
$m_\mathrm{Ni}$ 	&$0.505$/$0.636$	&$1.053$/$0.636$ 	&$0.688$/$0.636$	&$0.646$/$0.636$ \\
$E_\mathrm{ads}$ 	&$-159$			&$-98$			&$-41$			&$-55$ \\
$E_D - E_F$ 		&$-$				&$0.953$/$0.977$	&$0.572$/$0.565$	&$-0.266$/$-0.241$\\
$E_\mathrm{g}$ 	&$-$				&$51$/$74$		&$32$/$43$ 		&$62$/$51$\\
\hline
\end{tabular}

\end{table}

\clearpage

\clearpage

\clearpage

\clearpage

\end{document}